\documentclass[a4paper,11pt]{article}
\usepackage{pos}    
\usepackage{csquotes}   
\usepackage{subfig} 

\newcommand{\pt}{p_{\rm T}}
\newcommand{\ssqrt}{\sqrt{s}}

\newcommand{\Dplus}{{\rm D}^+}
\newcommand{\Dzero}{{\rm D}^0}
\newcommand{\Ds}{{\rm D^+_s}}
\newcommand{\Lc}{\Lambda^+_{\rm c}}
\newcommand{\SigmaC}{\Sigma_{\rm c}^{0,+,++}(2455)}
\newcommand{\XicZero}{\Xi_{\rm c}^0}
\newcommand{\XicPlus}{\Xi_{\rm c}^+}
\newcommand{\Xic}{\Xi_{\rm c}^{0,+}}
\newcommand{\OmegaC}{\Omega_{\rm c}^0}
\newcommand{\antiBzero}{\bar{\rm B}^0}
\newcommand{\Bminus}{{\rm B}^-}
\newcommand{\Bs}{{\rm B^0_s}}
\newcommand{\Lb}{\Lambda^0_{\rm b}}
\newcommand{\Bc}{{\rm B^-_c}}

\title{Heavy-flavour hadron production}

\author*[a]{Mattia Faggin}
\author{ (on behalf of the ALICE, ATLAS, CMS and LHCb Collaborations)}

\affiliation[a]{Universit\'a and INFN Padova,\\
  Dipartimento di Fisica e Astronomia \enquote{Galileo Galilei}, \\
  Via Marzolo 8, Padova, Italy}

\emailAdd{mattia.faggin@cern.ch}

\abstract{The conventional description of heavy-flavour hadron production in pp collisions is based on a factorisation approach, assuming universal fragmentation functions among collision systems. Recent results on heavy-flavour baryon measurements from the LHC experiments show tensions with model calculations based on this approach and employing fragmentation functions constrained from $\rm e^+e^-$ and $\rm e^-p$ collision experiments. In this contribution, the most recent results from ALICE, ATLAS, CMS and LHCb experiments on the heavy-flavour hadron production in pp collisions at the TeV scale are reported. The comparison with the theoretical predictions that address the baryon enhancement in hadronic collisions at the LHC is also discussed.}

\FullConference{%
 
 The Ninth Annual Conference on Large Hadron Collider Physics - LHCP2021\\
 7-12 June 2021\\
 Online
}

\begin{document}
\maketitle

\begin{figure}
    \centering
    \includegraphics[width=0.40\textwidth]{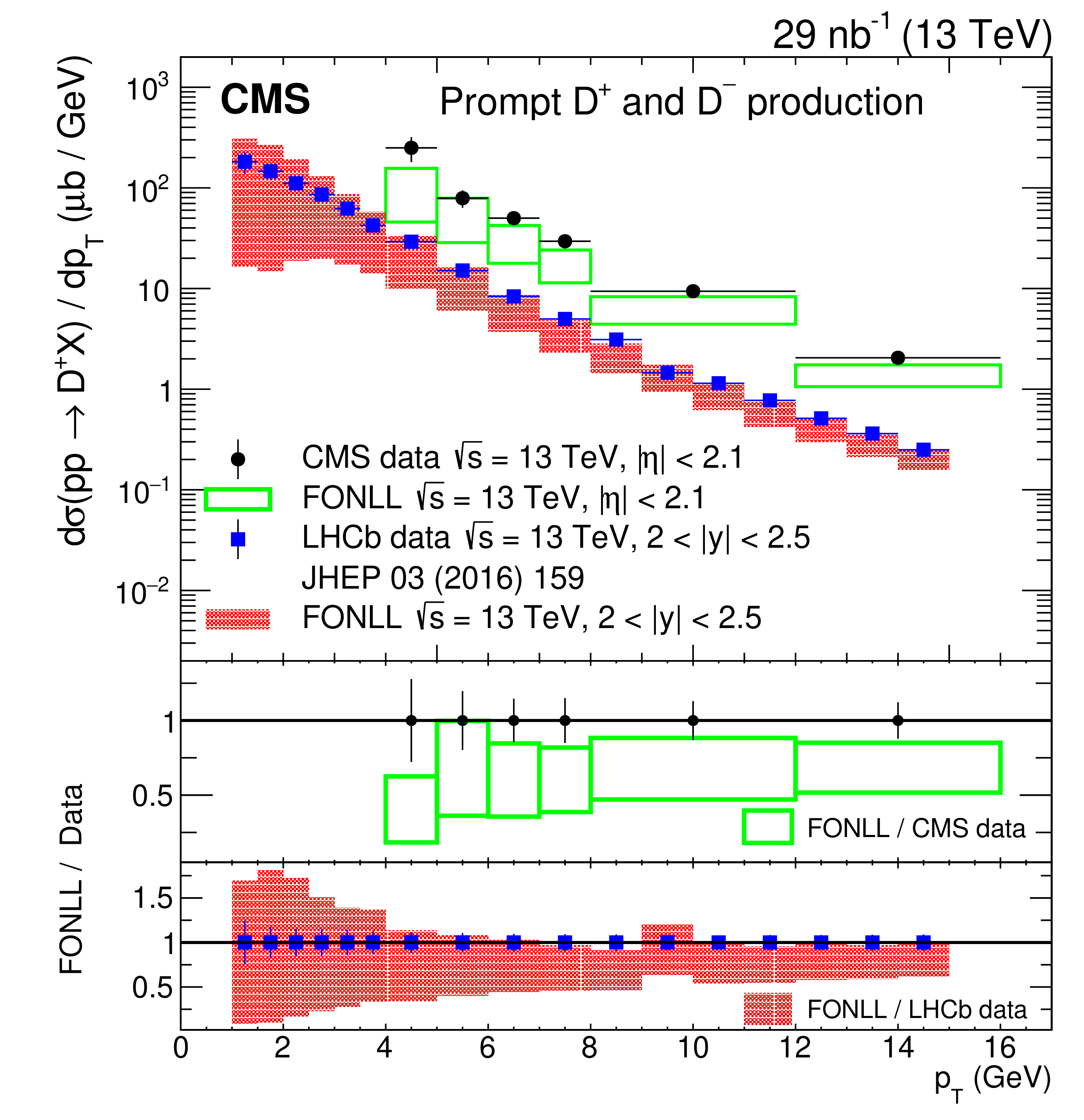} \quad
    \includegraphics[width=0.55\textwidth]{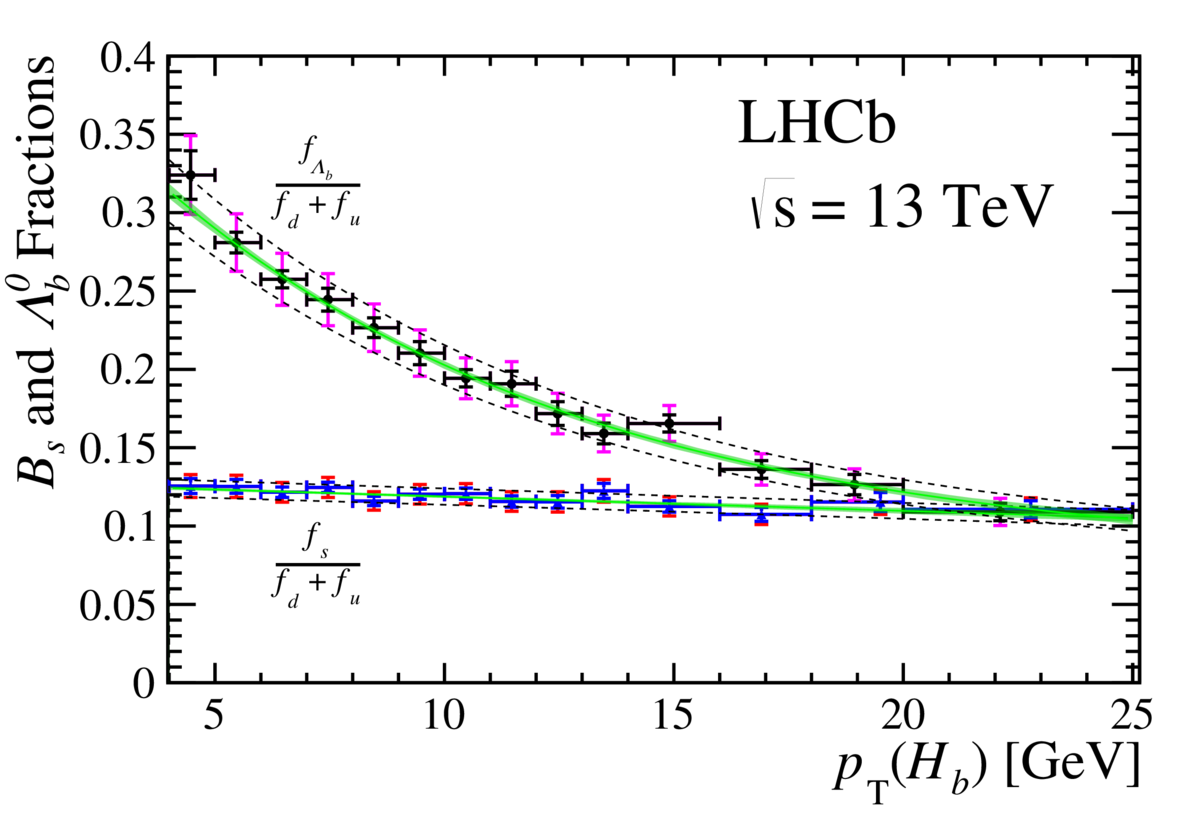} \\
    \caption{Left: production cross section of $\Dplus$ mesons at midrapidity ($|\eta|<2.1$) and forward rapidity ($2<y<2.5$) measured by the CMS \cite{CMS:2021lab} and LHCb \cite{LHCb:2016ikn} experiments respectively in pp collisions at $\ssqrt=13$ TeV, in comparison with FONLL predictions. Right: $\Bs/(\antiBzero+\Bminus)$ and $\Lb/(\antiBzero+\Bminus)$ ratios as a function of $\pt$ in pp collisions at $\ssqrt=13$ TeV measured by the LHCb experiment \cite{LHCb:2019fns}.}
    \label{fig:baryon_over_meson_1}
\end{figure}

\begin{figure}
    \centering
    \includegraphics[width=0.36\textwidth]{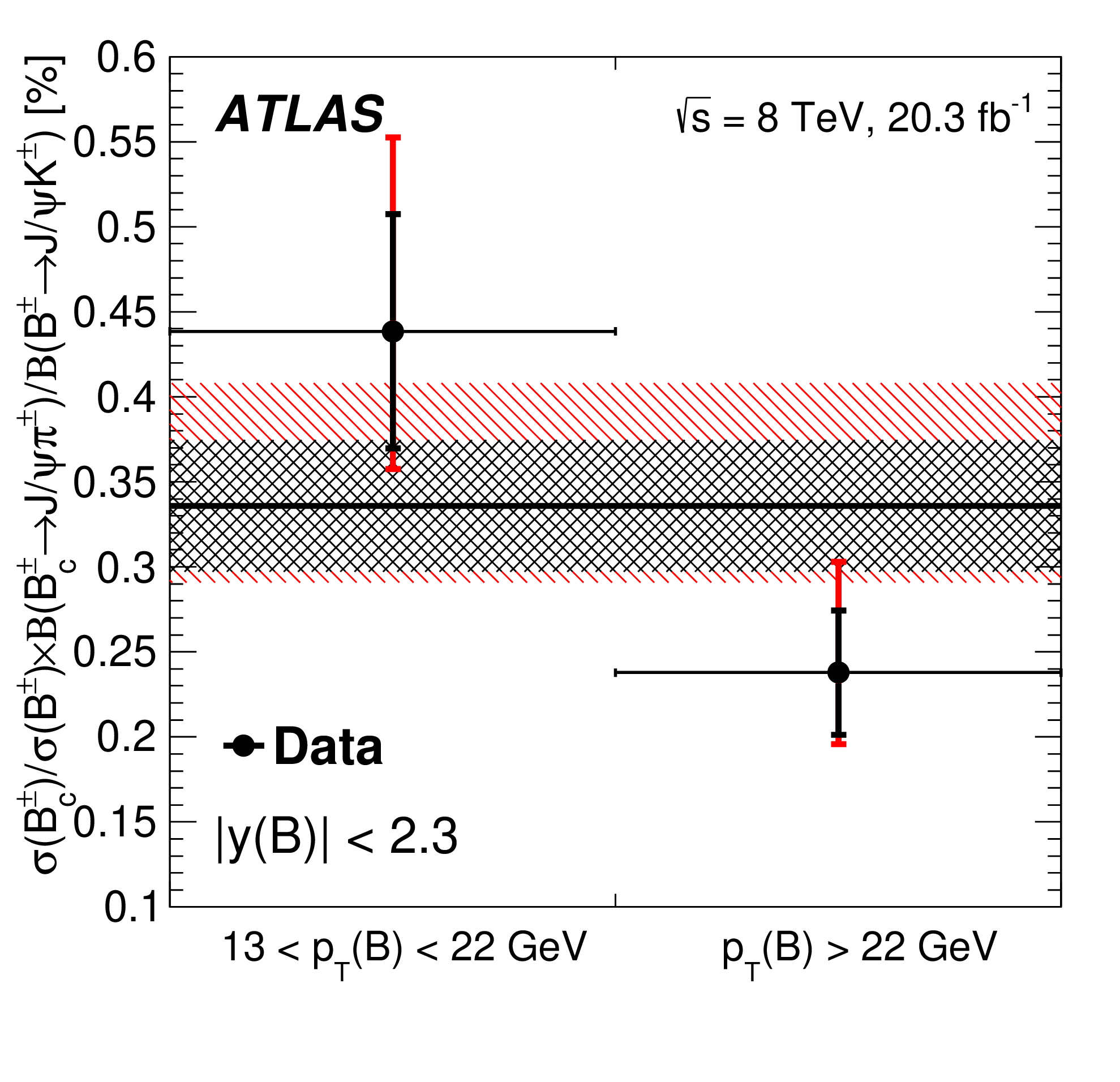} \quad
    \includegraphics[width=0.37\textwidth]{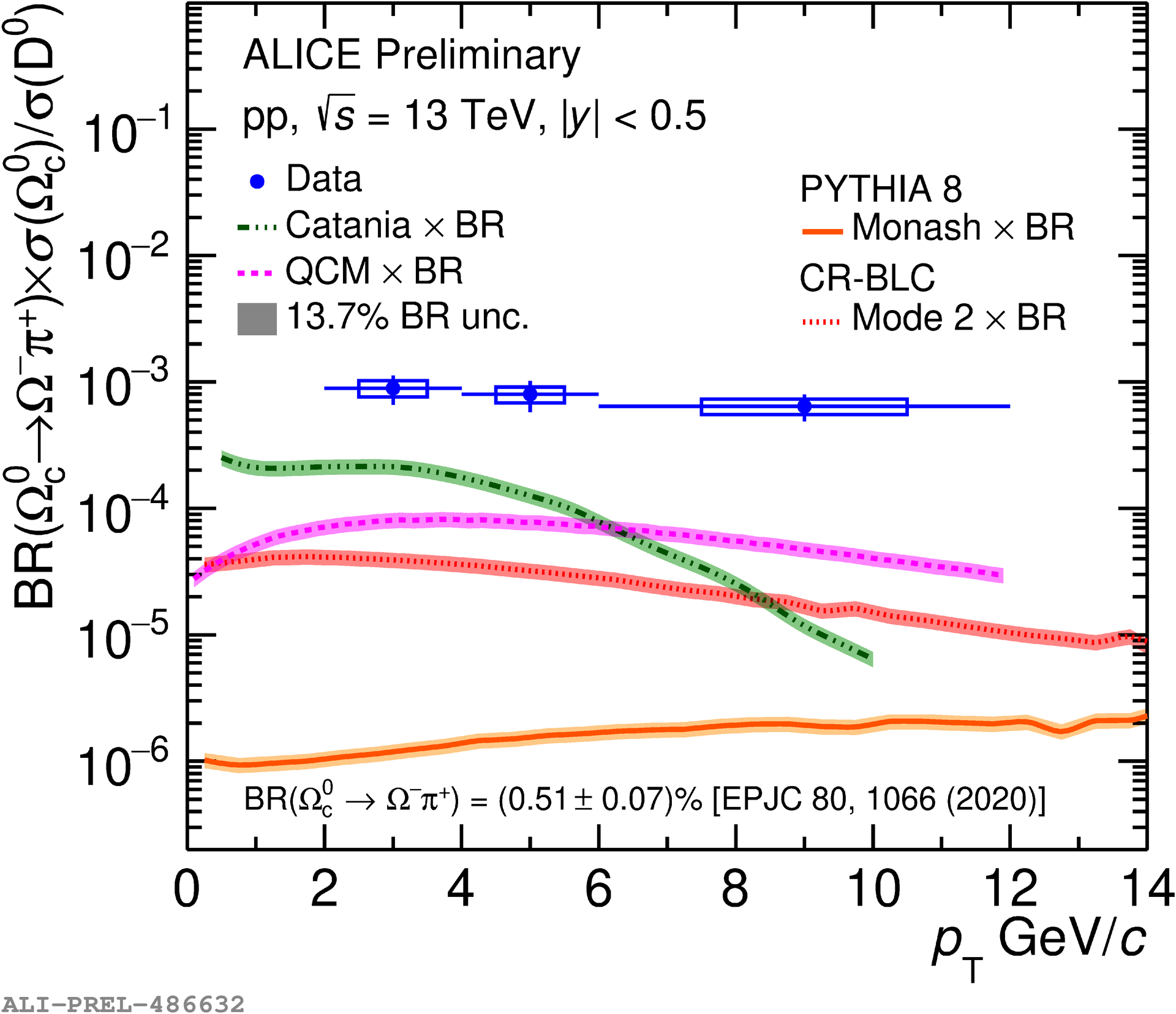} \\
    \caption{Left: ratio of $\sigma \times $BR between B$\rm ^\pm_c$ and B$^\pm$ as a function of transverse momentum at midrapidity ($|y(\rm B)|<2.3$) measured by the ATLAS experiment in pp collisions at $\ssqrt=8$ TeV \cite{ATLAS:2019jpi}. Right: $\OmegaC/\Dzero$ ratio measured in pp collisions at $\ssqrt=13$ TeV by the ALICE experiment, in comparison with different theoretical predictions.}
    \label{fig:baryon_over_meson_2}
\end{figure}

\noindent
The measurements of heavy-flavour hadron production at the LHC are fundamental tests of perturbative QCD calculations in proton-proton (pp) collisions. The standard description of heavy-flavour hadron production is based on a factorization approach, according to which it can be expressed as the convolution of: (a) the parton distribution functions (PDFs), describing how the partons are distributed within the colliding protons; (b) the cross section of the partonic scattering, producing the heavy quarks;
(c) the fragmentation functions, which quantify the probability for a quark to produce a hadron of a certain species with a given momentum fraction. The latter ingredients, usually considered universal among collision systems, are constrained from $\rm e^+e^-$ and $\rm e^-p$ collision measurements. The theoretical calculations based on this approach successfully describe the precise measurements of heavy-flavour meson production in pp collisions at the LHC \cite{ALICE:2021mgk,CMS:2018eso,CMS:2021mzx,ATLAS:2013cia,LHCb:2016ikn,CMS:2021lab} (left panel of Fig. \ref{fig:baryon_over_meson_1}).
The relative abundances of charm hadrons are sensitive to the charm quark hadronization and they provide a tool for the measurement of the charm fragmentation fraction ratios. The prompt $\Dplus/\Dzero$ and $\Ds/\Dzero$ ratios measured in pp collisions at mid and forward rapidity at the LHC are almost independent on the meson transverse momentum ($\pt$) and are in agreement with $\rm e^+e^-$ measurements, as well as with calculations relying on universal fragmentation functions \cite{ALICE:2021mgk,LHCb:2016ikn}. Similar observations are valid for the beauty mesons, for example, the $\Bs/(\antiBzero+\Bminus)$ ratio shown in the right panel of Fig. \ref{fig:baryon_over_meson_1} \cite{LHCb:2019fns}. The ratios between strange and non-strange mesons for the charm and beauty sectors are compatible with each other, and they do not show any strong dependence on the collision energy within uncertainties \cite{ALICE:2021mgk,LHCb:2021qbv}.
Instead, the baryon-to-meson ratio shows a clear $\pt$ dependence visible in the $\Lb/(\antiBzero+\Bminus)$ ratio measured by the LHCb experiment (right panel of Fig. \ref{fig:baryon_over_meson_1}), which is significantly enhanced compared to the $\Bs/(\antiBzero+\Bminus)$ ratio at low $\pt$. 
A proposed hypothesis is based on the mass difference among the hadrons involved 
($m_{\Lb}(\sim 5.6$ GeV/$c^2) > m_{\rm B}(\sim 5.3$ GeV/$c^2)$), where a coalescence-like process may enhance the production of heavier particles. Such an explanation is not supported by the $\Bc$ production in pp collisions at the LHC, as measured by the LHCb and ATLAS experiments \cite{LHCb:2019tea,ATLAS:2019jpi} (left panel of Fig. \ref{fig:baryon_over_meson_2}). The $\Bc/(\antiBzero+\Bminus)$ ratio shows a milder $\pt$ dependence with respect to the $\Lb/(\antiBzero+\Bminus)$ despite the larger mass of the $\Bc$ hadron ($m_{\Bc}(\sim 6.3$ GeV/$c^2) > m_{\Lb}(\sim 5.6$ GeV/$c^2)$). 
Recently, the CMS collaboration measured the $\rm B_c(2S)^+$ and $\rm B_c^*(2S)^+$ production in pp collisions at $\ssqrt=13$ TeV \cite{CMS:2020rcj}, which can provide useful insights in the ground-state $\Bc$ production. Their ratios compared to $\Bc$ do not show any significant dependence on $\pt$.
The baryon-to-meson ratio is studied also for the charm sector and a significant $\pt$ dependence is observed for the $\Lc/\Dzero$ ratio measured by the ALICE and CMS experiments in pp collisions at $\ssqrt=5.02, 7, 13$ TeV \cite{CMS:2019uws,Acharya:2017kfy,Acharya:2020lrg,Acharya:2020uqi,Acharya:2021vpo}. This ratio is higher by a factor of 2.5 to 5, depending on the $\pt$, to the results from $\rm e^+e^-$ and $\rm e^-p$ collisions, which are almost independent on the hadron $\pt$. The baryon-to-meson ratio enhancement observed at low $\pt$ may suggest that the fragmentation functions of the heavy-flavour sector are not universal. 
The baryon enhancement observed in pp collisions at the LHC was investigated also by the theory community. 
In the PYTHIA 8 event generator with improved colour reconnection mechanisms \cite{Christiansen:2015yqa} the baryon production is enhanced by the junction topology, a colour reconnection mechanism beyond the leading colour approximation, 
giving the possibility to partons from all the multi-parton interactions in the colliding protons and those from the beam remnants to combine. 
In the Quark (re-)Combination Mechanism (QCM \cite{Song:2018tpv}) model the charm hadron production at low $\pt$ is explained by the coalescence of charm quarks produced in hard scatterings with equal-velocity 
light quarks from fragmentation. This mechanism rules the charm hadron production at low $\pt$ also in the Catania model \cite{Minissale:2020bif}, where the hadronization is conceived as an interplay between fragmentation and coalescence with light quarks present in a thermalised system of u, d, s quarks and gluons. Another approach 
is based on the Statistical Hadronization Model (SHM), where the hadronization is ruled by thermo-statistical weights governed by the hadron masses, and on the presence of an augmented set of excited charm baryons predicted by the Relativistc Quark Model (RQM), whose strong decays increment the amount of ground-state charm baryons \cite{HE2019117}.

Precise measurements of charm hadron production are fundamental to understand the underlying mechanisms ruling the charm quark hadronization in pp collisions at the LHC. The ALICE experiment measured for the first time the $\SigmaC$ production in pp collisions at $\ssqrt=13$ TeV at midrapidity \cite{Acharya:2021vpo} where the ratio to $\Dzero$ production yield is shown to be significantly higher than those in $\rm e^+e^-$ and $\rm e^-p$ results, showing a larger relative enhancement compared to that of the $\Lc/\Dzero$ ratio. The $\SigmaC/\Dzero$ enhancement 
partially accounts for the $\Lc/\Dzero$ one and it is described within uncertainties by the model predictions mentioned above. The fraction of $\Lc$ baryons from strong decays of $\SigmaC$ states amounts to about 38\% in the range $2<\pt<12$ GeV/$c$, more than a factor 2 higher than $\rm e^+e^-$ and $\rm e^-p$ collisions. This result is overestimated by the PYTHIA 8 colour reconnection modes, suggesting that some ingredients are missing for the description of the direct $\Lc$ production. A significant enhancement with respect to the $\rm e^+e^-$ and $\rm e^-p$ results is observed also for the $\Xic/\Dzero$ ratio in pp collisions at $\ssqrt = 5.02, 13$ TeV \cite{Acharya:2021dsq,Acharya:2021vjp}, but in this case all the model predictions significantly underestimate the measurement. 
A similar comparison is observed for the $\OmegaC/\Dzero$ ratio, (right panel of Fig. \ref{fig:baryon_over_meson_2}), where the baryon-to-meson ratio measured by the ALICE experiment is significantly higher than the model predictions. 
This new result suggests that a sizeable contribution of charm hadronization in pp collisions at the LHC energies may be ascribed to the $\OmegaC$ production. Thanks to the measurement of the production cross section of all charm ground-state hadrons, the ALICE experiment measured the charm fragmentation fractions in pp collisions at $\ssqrt=5.02$ TeV \cite{ALICE:2021dhb}, as the cross section of each species normalised by the sum of those of all ground states. A decrease of a factor $1.2\div 1.4$ is observed for $\Dzero$ and $\Dplus$ mesons in pp collisions with respect to $\rm e^+e^-$ and $\rm e^-p$ collisions, accompanied by an increase of a factor 3 for the $\Lc$ fragmentation fraction and a $\XicZero$ and $\XicPlus$ production of the same order of $\Ds$ meson. Such a baryon enhancement observed at the LHC challenges the universality of fragmentation functions. The production cross section of each charm ground-state hadron enabled the measurement of the $\rm c\bar{c}$ production cross section in pp collisions at $\ssqrt=5.02$ TeV at midrapidity as the sum of the ground-state hadron cross sections and to improve the previously published results at $\ssqrt=2.76, 7$ TeV from D mesons with the measured fragmentation fractions. Due to the baryon enhancement, 
the $\rm c\bar{c}$ production cross sections at $\ssqrt = 2.76, 7$ TeV increased of about 40\%. The results at $\ssqrt=2.76, 5.02, 7$ TeV lie on the upper edge of FONLL \cite{Cacciari:2012ny} and NLO \cite{dEnterria:2016ids} calculations.

Further insights in the charm quark hadronization may come from the measurement of multi-charm baryons. A first measurement of the $\Xi_{\rm cc}^{++}$ double-charm baryon production relative to that of the $\Lc$ in pp collisions at $\ssqrt=13$ TeV is provided by the LHCb collaboration \cite{articleXicc}, but more differential measurements are necessary to further understand the mechanisms underlying the charm quark hadronization. Finally, the charm hadron production is studied as a function of event multiplicity via the measurement of the $\Lc/\Dzero$ in pp collisions at $\ssqrt=13$ TeV by the ALICE experiment. The measured ratio smoothly increases from pp to lead-lead (Pb--Pb) collisions in $2<\pt<8$ GeV/$c$, suggesting that similar processes may rule the charm hadronization in several collision systems. This conclusion is still under debate, due to flow effects that may be relevant in Pb--Pb collisions.

In conclusion, the recent results on heavy-flavour hadron production in pp collisions at the LHC from ALICE, ATLAS, CMS and LHCb experiments challenge the assumed universality of fragmentation functions, given an enhanced charm baryon production observed in pp collisions at the TeV scale. A joint effort between theory and experiments is required to understand this phenomenon and a complete comprehension of the mechanisms ruling the charm quark hadronization can benefit from the measurement of multi-charm baryon and quarkonium production.

%
%
\bibliographystyle{JHEP}
\bibliography{bibliography}

\providecommand{\href}[2]{#2}\begingroup\raggedright\begin{thebibliography}{10}

\bibitem{CMS:2021lab}
{\scshape CMS} collaboration, \emph{{Measurement of prompt open-charm
  production cross sections in proton-proton collisions at $\sqrt{s} = $ 13
  TeV}},  \href{https://arxiv.org/abs/2107.01476}{{\ttfamily 2107.01476}}.

\bibitem{LHCb:2016ikn}
{\scshape LHCb} collaboration, \emph{{Measurements of prompt charm production
  cross-sections in pp collisions at $ \sqrt{s}=5 $ TeV}},
  \href{https://doi.org/10.1007/JHEP06(2017)147}{\emph{JHEP} {\bfseries 06}
  (2017) 147} [\href{https://arxiv.org/abs/1610.02230}{{\ttfamily
  1610.02230}}].

\bibitem{LHCb:2019fns}
{\scshape LHCb} collaboration, \emph{{Measurement of $b$ hadron fractions in 13
  TeV $pp$ collisions}},
  \href{https://doi.org/10.1103/PhysRevD.100.031102}{\emph{Phys. Rev. D}
  {\bfseries 100} (2019) 031102}
  [\href{https://arxiv.org/abs/1902.06794}{{\ttfamily 1902.06794}}].

\bibitem{ALICE:2021mgk}
{\scshape ALICE} collaboration, \emph{{Measurement of beauty and charm
  production in pp collisions at $ \sqrt{s} $ = 5.02 TeV via non-prompt and
  prompt D mesons}}, \href{https://doi.org/10.1007/JHEP05(2021)220}{\emph{JHEP}
  {\bfseries 05} (2021) 220}
  [\href{https://arxiv.org/abs/2102.13601}{{\ttfamily 2102.13601}}].

\bibitem{CMS:2018eso}
{\scshape CMS} collaboration, \emph{{Measurement of B$^0_\mathrm{s}$ meson
  production in pp and PbPb collisions at $\sqrt{s_\mathrm{NN}} =$ 5.02 TeV}},
  \href{https://doi.org/10.1016/j.physletb.2019.07.014}{\emph{Phys. Lett. B}
  {\bfseries 796} (2019) 168}
  [\href{https://arxiv.org/abs/1810.03022}{{\ttfamily 1810.03022}}].

\bibitem{CMS:2021mzx}
{\scshape CMS} collaboration, \emph{{Observation of B$^0_s$ mesons and
  measurement of the B$^0_s$/B$^+$ yield ratio in PbPb collisions at
  $\sqrt{s_\mathrm{NN}}$ = 5.02 TeV}},
  \href{https://arxiv.org/abs/2109.01908}{{\ttfamily 2109.01908}}.

\bibitem{ATLAS:2013cia}
{\scshape ATLAS} collaboration, \emph{{Measurement of the differential
  cross-section of $B^{+}$ meson production in pp collisions at $\sqrt{s}$ = 7
  TeV at ATLAS}}, \href{https://doi.org/10.1007/JHEP10(2013)042}{\emph{JHEP}
  {\bfseries 10} (2013) 042} [\href{https://arxiv.org/abs/1307.0126}{{\ttfamily
  1307.0126}}].

\bibitem{LHCb:2021qbv}
{\scshape LHCb} collaboration, \emph{{Precise measurement of the $f_s/f_d$
  ratio of fragmentation fractions and of $B^0_s$ decay branching fractions}},
  \href{https://arxiv.org/abs/2103.06810}{{\ttfamily 2103.06810}}.

\bibitem{LHCb:2019tea}
{\scshape LHCb} collaboration, \emph{{Measurement of the $B_c^-$ meson
  production fraction and asymmetry in 7 and 13 TeV $pp$ collisions}},
  \href{https://doi.org/10.1103/PhysRevD.100.112006}{\emph{Phys. Rev. D}
  {\bfseries 100} (2019) 112006}
  [\href{https://arxiv.org/abs/1910.13404}{{\ttfamily 1910.13404}}].

\bibitem{ATLAS:2019jpi}
{\scshape ATLAS} collaboration, \emph{{Measurement of the relative
  $B^{\pm}_{c}/B^{\pm}$ production cross section with the ATLAS detector at
  $\sqrt{s}=8$ TeV}},
  \href{https://doi.org/10.1103/PhysRevD.104.012010}{\emph{Phys. Rev. D}
  {\bfseries 104} (2021) 012010}
  [\href{https://arxiv.org/abs/1912.02672}{{\ttfamily 1912.02672}}].

\bibitem{CMS:2020rcj}
{\scshape CMS} collaboration, \emph{{Measurement of B$_\mathrm{c}$(2S)$^+$ and
  B$_\mathrm{c}^*$(2S)$^+$ cross section ratios in proton-proton collisions at
  $\sqrt{s} =$ 13 TeV}},
  \href{https://doi.org/10.1103/PhysRevD.102.092007}{\emph{Phys. Rev. D}
  {\bfseries 102} (2020) 092007}
  [\href{https://arxiv.org/abs/2008.08629}{{\ttfamily 2008.08629}}].

\bibitem{CMS:2019uws}
{\scshape CMS} collaboration, \emph{{Production of $\Lambda_\mathrm{c}^+$
  baryons in proton-proton and lead-lead collisions at $\sqrt{s_\mathrm{NN}}=$
  5.02 TeV}}, \href{https://doi.org/10.1016/j.physletb.2020.135328}{\emph{Phys.
  Lett. B} {\bfseries 803} (2020) 135328}
  [\href{https://arxiv.org/abs/1906.03322}{{\ttfamily 1906.03322}}].

\bibitem{Acharya:2017kfy}
{\scshape ALICE} collaboration, \emph{{$\Lambda_{\rm c}^+$ production in pp
  collisions at $\sqrt{s} = 7$ TeV and in p-Pb collisions at $\sqrt{s_{\rm NN}}
  = 5.02$ TeV}}, \href{https://doi.org/10.1007/JHEP04(2018)108}{\emph{JHEP}
  {\bfseries 04} (2018) 108}
  [\href{https://arxiv.org/abs/1712.09581}{{\ttfamily 1712.09581}}].

\bibitem{Acharya:2020lrg}
{\scshape ALICE} collaboration, \emph{{$\Lambda_{\rm c}^{+}$ production in pp
  and in p-Pb collisions at $\sqrt{s_{\rm {NN}}} = 5.02$ TeV}},
  \href{https://arxiv.org/abs/2011.06079}{{\ttfamily 2011.06079}}.

\bibitem{Acharya:2020uqi}
{\scshape ALICE} collaboration, \emph{{$\Lambda_{\rm c}^{+}$ production and
  baryon-to-meson ratios in pp and p-Pb collisions at $\sqrt{s_\mathrm{NN}} =
  5.02$ TeV at the LHC}},  \href{https://arxiv.org/abs/2011.06078}{{\ttfamily
  2011.06078}}.

\bibitem{Acharya:2021vpo}
{\scshape ALICE} collaboration, \emph{{Measurement of prompt D$^{0}$,
  $\Lambda_{c}^{+}$, and $\Sigma_{c}^{0,++}$(2455) production in pp collisions
  at $\sqrt{s}$ = 13 TeV}},  \href{https://arxiv.org/abs/2106.08278}{{\ttfamily
  2106.08278}}.

\bibitem{Christiansen:2015yqa}
J.R.~Christiansen and P.Z.~Skands, \emph{{String Formation Beyond Leading
  Colour}}, \href{https://doi.org/10.1007/JHEP08(2015)003}{\emph{JHEP}
  {\bfseries 08} (2015) 003}
  [\href{https://arxiv.org/abs/1505.01681}{{\ttfamily 1505.01681}}].

\bibitem{Song:2018tpv}
J.~Song, H.-h.~Li and F.-l.~Shao, \emph{{New feature of low $p_{T}$ charm quark
  hadronization in $pp$ collisions at $\sqrt{s}=7$ TeV}},
  \href{https://doi.org/10.1140/epjc/s10052-018-5817-x}{\emph{Eur. Phys. J. C}
  {\bfseries 78} (2018) 344}
  [\href{https://arxiv.org/abs/1801.09402}{{\ttfamily 1801.09402}}].

\bibitem{Minissale:2020bif}
V.~Minissale, S.~Plumari and V.~Greco, \emph{{Charm Hadrons in pp collisions at
  LHC energy within a Coalescence plus Fragmentation approach}},
  \href{https://arxiv.org/abs/2012.12001}{{\ttfamily 2012.12001}}.

\bibitem{HE2019117}
M.~He and R.~Rapp, \emph{Charm-baryon production in proton-proton collisions},
  \href{https://doi.org/https://doi.org/10.1016/j.physletb.2019.06.004}{\emph{Physics
  Letters B} {\bfseries 795} (2019) 117}.

\bibitem{Acharya:2021dsq}
{\scshape ALICE} collaboration, \emph{{Measurement of the production cross
  section of prompt $\Xi^0_{\rm c}$ baryons at midrapidity in pp collisions at
  $\sqrt{s}$ = 5.02 TeV}},  \href{https://arxiv.org/abs/2105.05616}{{\ttfamily
  2105.05616}}.

\bibitem{Acharya:2021vjp}
{\scshape ALICE} collaboration, \emph{{Measurement of the cross sections of
  $\Xi^0_{\rm c}$ and $\Xi^+_{\rm c}$ baryons and branching-fraction ratio
  BR($\Xi^0_{\rm c} \rightarrow \Xi^-{\rm e}^+\nu_{\rm e}$)/BR($\Xi^0_{\rm c}
  \rightarrow \Xi^-\pi^+$) in pp collisions at 13 TeV}},
  \href{https://arxiv.org/abs/2105.05187}{{\ttfamily 2105.05187}}.

\bibitem{ALICE:2021dhb}
{\scshape ALICE} collaboration, \emph{{Charm-quark fragmentation fractions and
  production cross section at midrapidity in pp collisions at the LHC}},
  \href{https://arxiv.org/abs/2105.06335}{{\ttfamily 2105.06335}}.

\bibitem{Cacciari:2012ny}
M.~Cacciari, S.~Frixione, N.~Houdeau, M.L.~Mangano, P.~Nason and G.~Ridolfi,
  \emph{{Theoretical predictions for charm and bottom production at the LHC}},
  \href{https://doi.org/10.1007/JHEP10(2012)137}{\emph{JHEP} {\bfseries 10}
  (2012) 137} [\href{https://arxiv.org/abs/1205.6344}{{\ttfamily 1205.6344}}].

\bibitem{dEnterria:2016ids}
D.~d'Enterria and A.M.~Snigirev, \emph{{Triple parton scatterings in
  high-energy proton-proton collisions}},
  \href{https://doi.org/10.1103/PhysRevLett.118.122001}{\emph{Phys. Rev. Lett.}
  {\bfseries 118} (2017) 122001}
  [\href{https://arxiv.org/abs/1612.05582}{{\ttfamily 1612.05582}}].

\bibitem{articleXicc}
{\scshape LHCb} collaboration, \emph{Measurement of ${{\varXi_{cc}^{++}}}$
  production in pp collisions at ${\sqrt{ s}=13}$ tev},
  \href{https://doi.org/10.1088/1674-1137/44/2/022001}{\emph{Chinese Physics C}
  {\bfseries 44} (2020) 022001}.

\end{thebibliography}\endgroup

\end{document}